\documentclass[twocolumn]{aastex701}

\received{June 03, 2024} 
\revised{XXXX XX, XXXX} 
\accepted{XXXX XX, XXXX} 

\newcommand{\p}[1]{{\color{magenta}{#1}}}

\shorttitle{Coronal green-line Emission During a Prominence Eruption}
\shortauthors{Vemareddy et al}
\begin{document}
\title{Coronal green-line Emission During a Solar Prominence Eruption: Spectroscopic Evidence of Its Correspondence with the EUV Late Phase}
%
%
\author[orcid=0000-0003-4433-8823,sname='P']{P. Vemareddy} 
\affiliation{Indian Institute of Astrophysics, II Block, Koramangala, Bengaluru-560 034, India}
\email[show]{vemareddy@iiap.res.in} 
\author[orcid=0000-0003-4433-8823,sname='R']{R. Ramesh} 
\affiliation{Indian Institute of Astrophysics, II Block, Koramangala, Bengaluru-560 034, India}
\email{ramesh@iiap.res.in} 

\author[orcid=0000-0001-6093-3302,sname='M']{V. Muthu Priyal} 
\affiliation{Indian Institute of Astrophysics, II Block, Koramangala, Bengaluru-560 034, India}
\email{muthu.priyal@iiap.res.in} 
\begin{abstract}
We report spectroscopic observations during a prominence eruption on 9 October 2025 from the Sun’s west limb, associated with an M2.0 flare that commenced at 12:11 UT. The event was captured in sit-and-stare mode by the Visible Emission Line Coronagraph (VELC) onboard Aditya-L1, providing continuous observations of the Fe XIV 5303 \AA\ coronal greenline. The spatially averaged green-line emission exhibits a clear correspondence with the EUV late-phase (ELP) emission observed in multiple SDO/AIA channels. In particular, an enhanced 5303 \AA\ emission peak at 15:24 UT coincides with secondary peaks in the AIA 211, 171, 193, and 335 \AA\ passbands, occurring $\sim173$ minutes after the X-ray peak. The 5303 \AA\ line-width information provides important constraints on the mechanism responsible for the ELP. The line-width decreases progressively by about 28\% toward the ELP peak, broadly consistent with the cooling of post-flare coronal loops. This behavior suggests that the ELP is predominantly associated with plasma cooling, with relatively weak additional energy input from magnetic reconnection, which does not produce a significant enhancement in line-widths. The estimated non-thermal velocity decreases from $\sim$36 km s$^{-1}$ during the impulsive phase to $\sim$21 km s$^{-1}$ during the gradual phase, indicating a reduction in turbulent plasma motions. And doppler velocities evolve from predominantly positive values during the eruptive rise phase to negative values during the later gradual-phase, implying upward and subsequent downward plasma motions. To our knowledge, this study provides the first space-based spectroscopic evidence linking coronal green-line emission with the ELP and its origin during a solar flare. 
\end{abstract}

\section{Introduction}

Solar prominences are bright, structured features in the corona that can erupt and evolve into a coronal mass  (CME). When observed against the bright solar disk, they appear as dark filaments due to absorption (e.g., \citealt{TandbergHanssen1998,Marque2004_RadioObs_prom, MartinSaraF1998, Parenti2012, Vemareddy2012_FilErup, Gibson2018_SolProm_Theory_Obs, Vemareddy2017_PromEru, Wood2016_Prominence_1AU, Vemareddy2024_PromEru}). These features quite often erupt due to loss of equilibrium leading to strong CME which have the potential to propagate and impact near-earth environment (e.g., \citealt{Vemareddy2024_FilEru})). Understanding the formation and eruption mechanisms of these structures is essential for space-weather prediction, and they are therefore extensively studied using H$\alpha$, extreme ultraviolet (EUV), and white-light coronagraph observations.  

Spectroscopic observations of the coronal emission line at Fe XIV 5303~\AA~(the so-called green line),  first discovered from solar eclipse observations \citep{Young1870_Eclipse}, have long been employed in coronagraphic studies of the solar corona (e.g., \citealt{Arnaud1982_ObsPolaris_5303, Rybansky1994,JSingh1999_Spec_SpatialVar,WangYM1997_GreenLineCor_PhotMag, Habbal2010_HotPromShrouds,Singh2011_2009Ecl,Ding2017_FirstDet_PromMat,Koutchmy2019_NewDeepSpec,BenjaminBoe2020_CME_Induced_Thermodyn, Muro2023_VisEmiSpec_2019Ecl}). The emissivity is due to a forbidden transition process of the Fe XIV ion at 5303 \AA. Owing to its high formation temperature (1.8-2 MK) and strong emissivity this line serves as an effective diagnostic of warm coronal structures. In particular, green-line spectroscopy provides not only intensity measurements but also key information on line width and Doppler shift. The line width offers constraints on coronal heating processes, while Doppler shifts enable the inference of plasma motions along the line of sight (LoS). Together with multi-wavelength observations, the coronal green line observations could be used to deduce the parameters like plasma density, temperature, velocity, magnetic field and subsequently exploited for studies of coronal physics. 

Most spectroscopic studies of coronal emission lines have traditionally relied on ground-based observations, primarily aimed at diagnosing coronal properties such as density and temperature variations in different coronal structures. However, these studies have often been limited by the short observing windows available during solar eclipses and by atmospheric seeing constraints affecting ground-based telescopes. Using coronal green-line spectra obtained at the Norikura Solar Observatory, \citet{Sakurai2002_CoroWaves} investigated the presence of different types of coronal waves and their potential role in energy transport within the corona. 

Only a limited number of studies have focused on prominence structures and their eruptive dynamics using emission-line spectroscopy. By combining broadband white-light and narrow band emission-line observations acquired during the total solar eclipses of 29 March 2006 and 1 August 2008, \citet{Habbal2010_HotPromShrouds} demonstrated that prominences observed above the solar limb are embedded within hot plasma confined by twisted magnetic structures. In a subsequent study based on comprehensive spectral observations obtained during the total solar eclipse of 20 March 2015, \citet{Ding2017_FirstDet_PromMat} reported the unambiguous detection of a filamentary CME front at temperatures of approximately $2\times10^6$ K, containing inclusions of cooler prominence material. 

The recent launch of the Visible Emission Line Coronagraph (VELC) aboard Aditya-L1 has enabled routine spectroscopic observations in the coronal green line, significantly improving the opportunity to investigate transient eruptive events and their thermal evolution. Using VELC sit-and-stare observations, \citet{RameshR2024_NewRes} reported the onset of a CME on 16 July 2024 associated with an X1.9-class soft X-ray flare. Their analysis revealed an approximately 50\% decrease in coronal intensity near the CME source region, attributed to plasma depletion, along with an enhancement of nearly 15\% in line width. Similarly, \citet{MPriyal2025_NearSun} presented VELC observations of a flare-less CME that exhibited an increase of about 57\% in line intensity accompanied by a corresponding decrease of roughly 10\% in line width. 

In this study, we present VELC green line observations obtained during prominence eruption (PE) on October 9, 2025. Our primary focus is on the evolution of coronal plasma properties during the PE, which is driven by magnetic reconnection associated with the accompanying flare. In particular, we investigate how the coronal greenline emission evolves in relation to the EUV emission, to understand their temporal correspondence throughout the event. The Fe XIV 5303 \AA\ green line has a formation temperature closely matches with the plasma sampled by the AIA 211 \AA\ channel. In addition to EUV images, green-line spectroscopy provides valuable diagnostics, including Doppler velocities and line widths, which offer insights into the plasma motions and the thermal and non-thermal evolution of the coronal plasma. This paper is organized as follows: Section~\ref{AIAObs} presents the EUV imaging observations of the PE, section~\ref{VELCObs} analyses the VELC spectral data for the evolution of peak line emission, line width, doppler velocity during the PE, and a differential emission measure analysis is supplemented in section~\ref{DemAna} for the plasma temperature and finally summary with a discussion is given in Section~\ref{Summ}.
\begin{figure*}[!ht]
\centering
\includegraphics[width=0.99\linewidth]{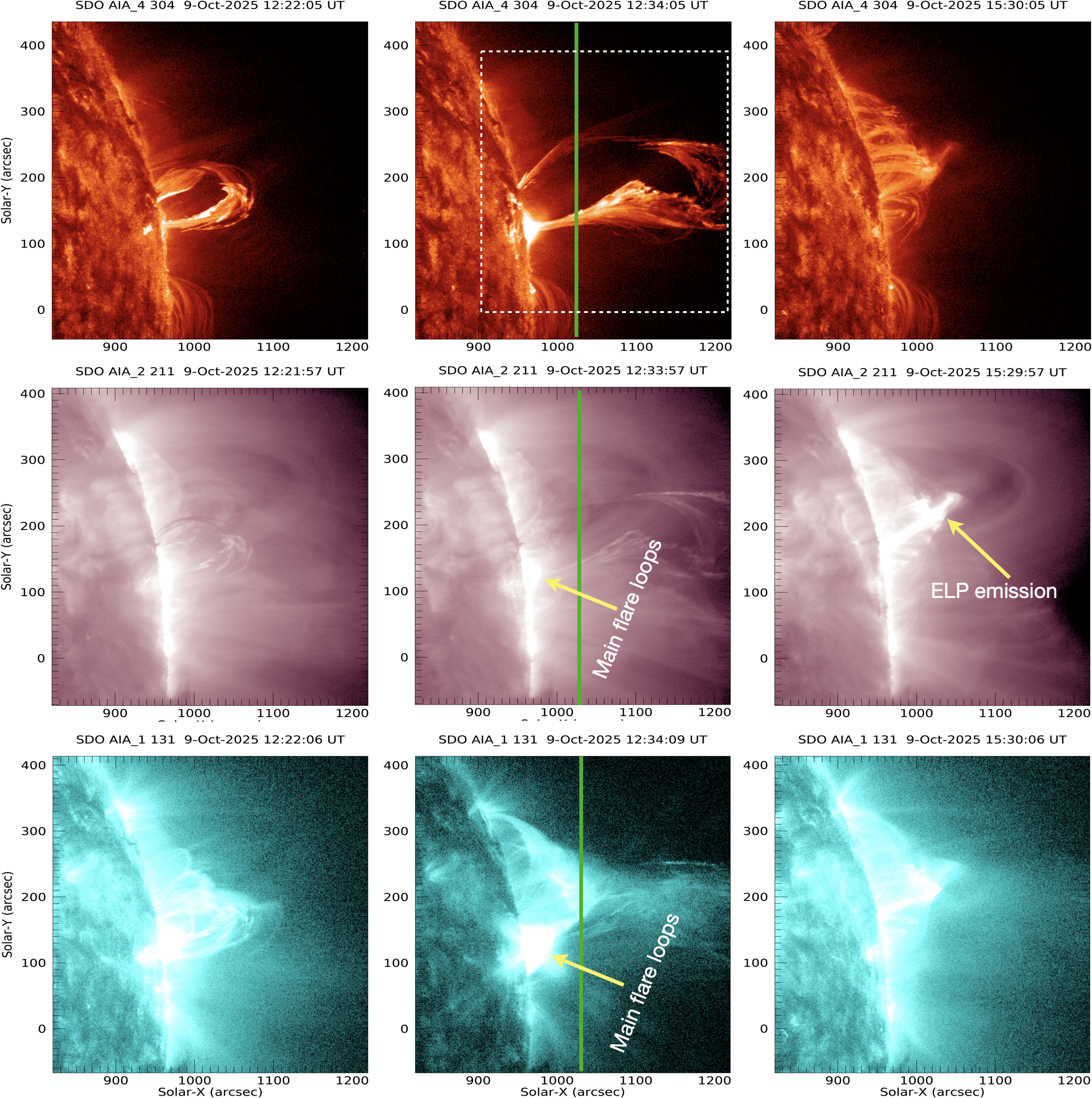}
\caption{AIA observations of the prominence eruption at the west limb are illustrated. {\bf Top row:} snapshots in AIA 304  \AA~capturing the slow-rise phase and subsequent post-eruption evolution in the corona. Dashed rectangular box refers to the region to derive lightcurves. {\bf Middle row:} AIA 211 \AA~images highlighting the prominence structure along with the surrounding coronal loop system. {\bf Bottom row:} AIA 131 \AA~images revealing hot flare plasma produced by magnetic reconnection beneath the rising prominence. The vertical green line at $X=1024$ arcsec marks the position of Slit 4 used for spectroscopic measurements with the VELC instrument on board Aditya-L1. (An animation of this figure is available online) }
\label{fig1}
\end{figure*}
\begin{figure}[!ht]
    \centering
    \includegraphics[width=0.99\linewidth]{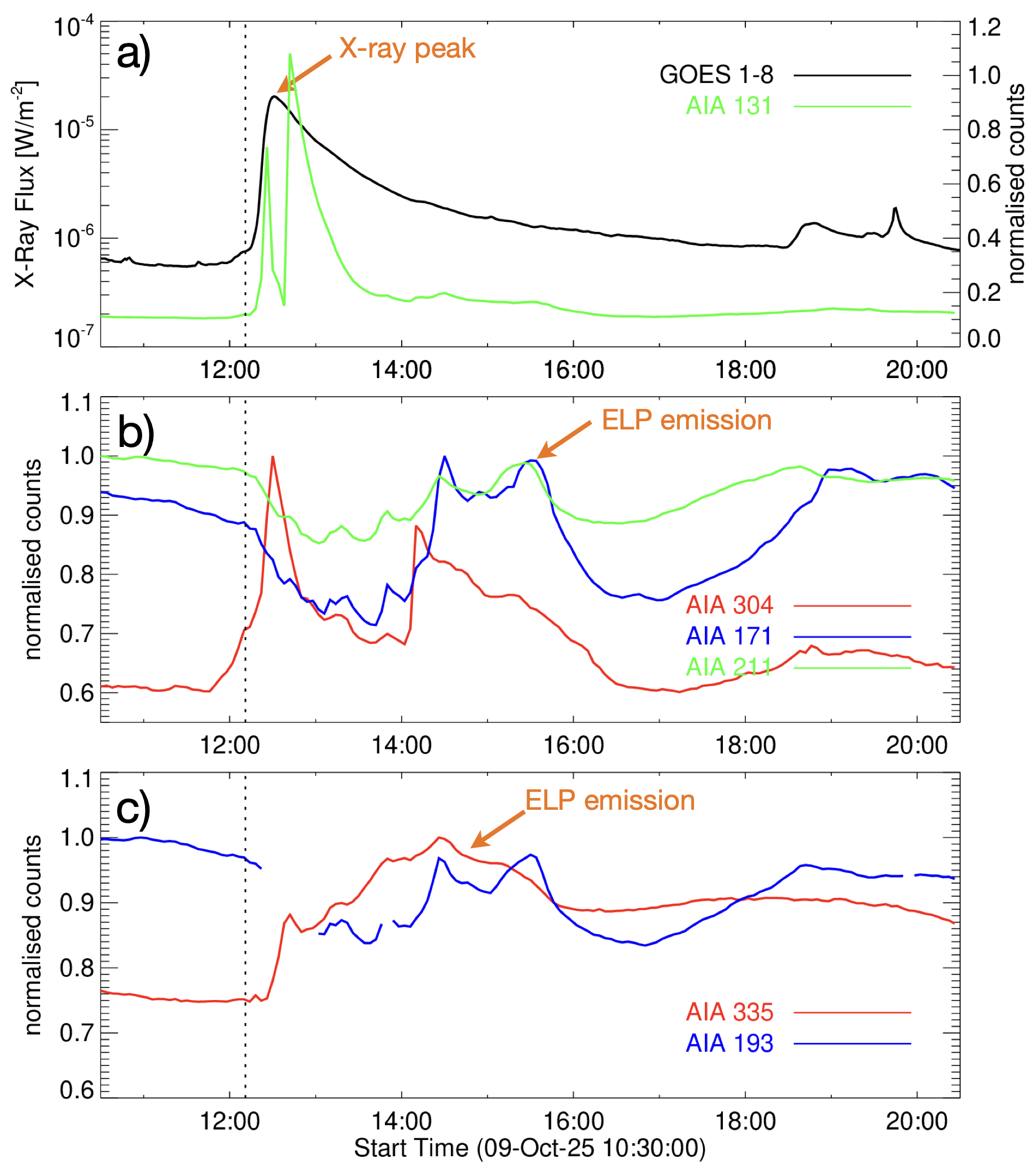}
    \caption{Light curves during PE and associated flare. {\bf a)} Light curves of {\bf GOES X-ray flux in 1-8~\AA\ band}~and AIA 131 \AA~flux as a function of time. {\bf b)} AIA 304, 171, 211 \AA\ light curves. {\bf c)} AIA 193, 335~\AA~light curves. The peak of X-ray flux corresponds to M2.0 flare and the vertical dotted line at 12:11 UT refers to flare start time. The X-ray flux peak is co-temporal with the 131 and 304~\AA~emissions; however, the gradual phase of the flare is accompanied by enhanced EUV emission with the peaks at 14:25 UT (335 \AA), 15:24 UT (171, 211, 193 \AA). }
    \label{fig_lgtcrv}
\end{figure}
\section{Prominence Eruption On 9 October 2025}
\label{AIAObs}
The prominence examined in this study is located on the western limb of the Sun on October 9, 2025. It is clearly observed in extreme ultraviolet (EUV) images from the Atmospheric Imaging Assembly (AIA; \citealt{lemen2012}) onboard the Solar Dynamics Observatory (SDO). The prominence originated from active region (AR) 14232, positioned at N02W91, and its eruption onset is illustrated in Figure~\ref{fig1}. In the AIA 304~\AA~channel, the prominence appears as a bright structure against the background, while the 211~\AA~channel highlights the surrounding coronal loop system. The 131~\AA~channel, sensitive to hot plasma (peak temperature response at $\approx$10 MK), primarily captures flare-related emission rather than the relatively cool magnetically dominated prominence structure.

Initially located at a height of approximately 30 Mm, the prominence underwent a slow-rise motion beginning at around 12:10 UT. By 12:34 UT, it had started to move beyond the AIA field of view (1.27~$R_\odot$), as evidenced by the 304 and 211~\AA~observations shown in Figure~\ref{fig1}. This phase is typically associated with magnetic reconnection beneath the rising prominence \citep{Vemareddy2012_FilErup}, leading to enhanced hot plasma emission visible in the AIA 131~\AA~channel. The slow upward motion of prominence evolves to an eventual eruption leading to a narrow CME (angular width of 51 degrees) as observed in Large Angle and Spectrometric Coronagraph (LASCO; \citealt{Brueckner1995_LASCO}) onboard Solar and Heliospheric Observatory (SoHO) white light observations and has a linear speed of 260 km\,s$^{-1}$.

Figure~\ref{fig_lgtcrv} presents the EUV light curves during the PE across multiple wavelengths. These light curves are derived by averaging the emission within the field of view (rectangular box) shown in Figure~\ref{fig1}. With the onset of the slow upward motion, magnetic reconnection sets in, leading to an increase in the GOES X-ray flux from 12:11 UT, marking the flare onset. This is followed by peak reconnection at 12:31 UT, corresponding to the maximum flare intensity, classified as an M2.0 event (\texttt{SOL2025-10-09T12:11}). The AIA 131 and 304~\AA~light curves closely track the X-ray flare profile, whereas the AIA 171, 193 and 211~\AA~channels exhibit a decrease in emission. This reduction is indicative of EUV dimming \citep{Thompson1998_CorDimm, Zarro1999_EIT_Dimm}, which results from plasma depletion during the CME eruption and the subsequent expansion of coronal loops, leading to lower density. The AIA 131\AA\ passband captures emission from hot plasma similar to X-ray because it has contribution from Fe XXI with peak temperature response at 10-11 MK \citep{lemen2012}. 

The main flare phase is followed by a gradual phase with enhanced EUV emission characteristic of the EUV late phase (ELP; \citealt{Woods2011_EVEObs_Flares, DaiY2013_EUV_LatePhase, LiuK2013_ELP}). During this phase, the AIA 304 \AA\ ($T\sim1.6$ MK) and 335 \AA\ ($T\sim2.5$ MK) light curves exhibit secondary peaks at 14:10 UT and 14:26 UT, respectively, occurring 99 and 115 minutes after the X-ray peak, without a corresponding enhancement in the hotter 131 \AA\ emission. From approximately 14:00 UT onward, the AIA 171 \AA\ ($T\sim0.6$ MK), 211 \AA\ ($T\sim2$ MK), and 193 \AA\ ($T\sim1.6$ MK) channels also show enhanced emission, with prominent peaks around 14:25 UT and 15:25 UT. The secondary enhancement with multiple peaks with relatively weak energy input, was suggested to successive episodes of magnetic reconnection \citep{DaiY2013_EUV_LatePhase}.

The ELP refers to a secondary enhancement of warm coronal EUV emission, typically at temperatures of $\sim1.5$--3 MK, occurring tens of minutes to several hours after the soft X-ray peak \citep{Woods2011_EVEObs_Flares,ZhongY2021_EUV_LatePhase} as this event. Such delayed EUV enhancements have been reported in both confined and eruptive flares \citep{Vemareddy2012_FilErup,ZhongY2021_EUV_LatePhase,DaiY2013_EUV_LatePhase}. In the present off-limb eruptive event, the AIA 211 and 131 \AA\ images (Figure~\ref{fig1}) indicate that the late-phase emission originates from a system of longer and higher-lying loops compared with the compact main flare loops. The earlier occurrence of the AIA 335 \AA\ peak, compared to the 211, 171, and 193 \AA\ channels, is consistent with the progressive cooling of plasma in these extended loops \citep{DaiY2013_EUV_LatePhase}. Notably, the absence of a comparable ELP signature in flare-less prominence eruptions \citep{Vemareddy2017_PromEru} further suggests that the observed late-phase emission is closely associated with flare-related heating and the subsequent long-lasting cooling of post-flare coronal plasma.



\begin{figure*}[!ht]
    \centering
    \includegraphics[width=0.99\linewidth]{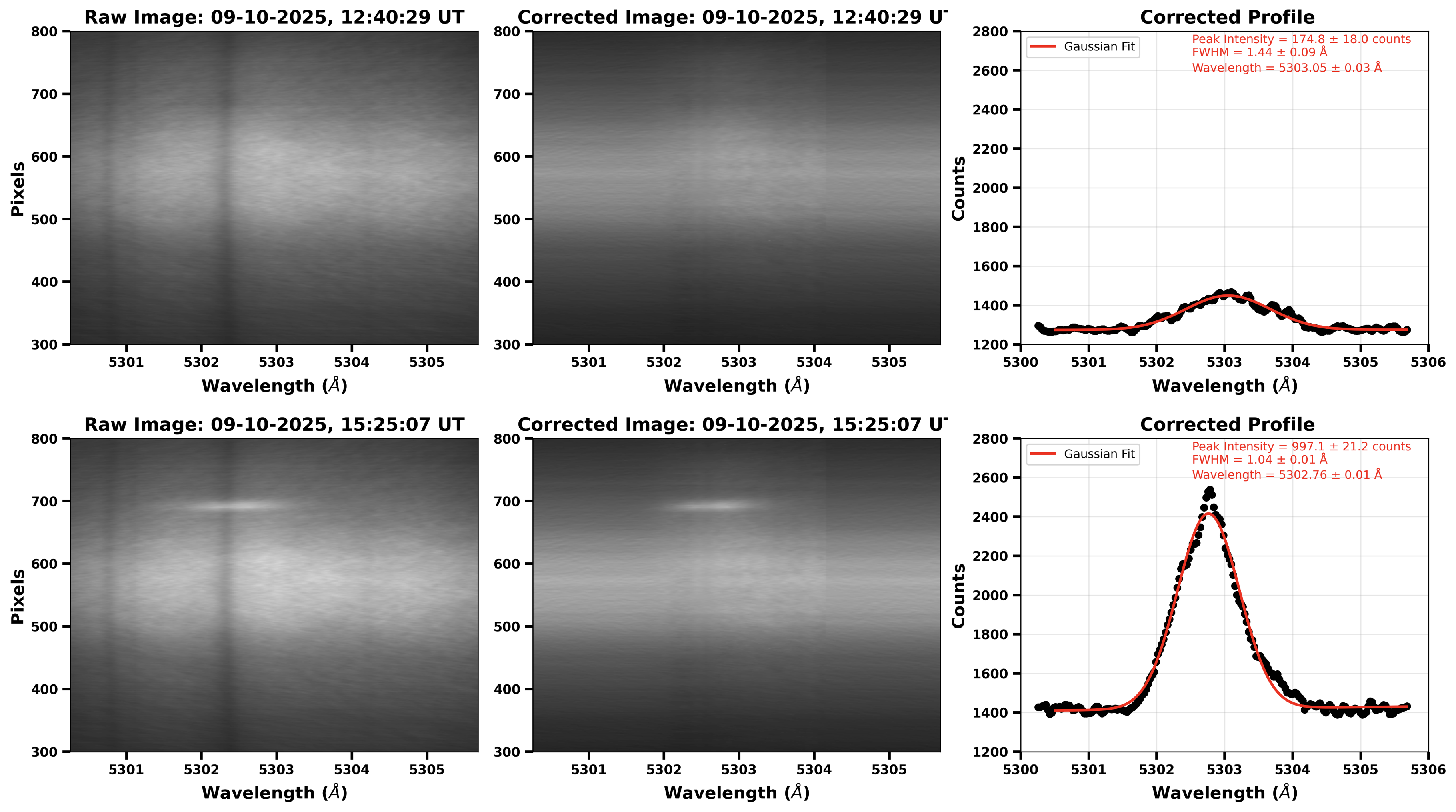}
    \caption{ {\bf left panels:} raw image of 5303 \AA~spectrum (slit 4) at 12:40 UT and 15:25 UT respectively. {\bf middle panels:} corrected image of spectrum after removing detector response (dark current \& flat-fielding) and absorption line, {\bf right panels:} extracted spectral profile at $Y=697$ pixel with Gaussian fit (red) . Derived parameters are annotated.}
    \label{fig_spec_prof}
\end{figure*}

\section{Sit and stare Spectroscopic observation by VELC}
\label{VELCObs}
Spectroscopic observations during the PE in the 5303~\AA~channel were obtained using the VELC aboard Aditya-L1 (observation ID: \texttt{VS1\_T25\_1473\_002796}), providing a unique opportunity to investigate the thermal properties of the erupting prominence. The observations employed four slits operating in both raster-scan and sit-and-stare modes, with a spatial sampling of 1.25 arcsec per pixel along the slit and a slit width 50 $\mu m$ that is equivalent to 9.6 arcsec. In the sit-and-stare mode, Slit 4 was positioned at approximately 1024 arcsec ($\approx1.06~R_\odot$) on the west limb, capturing the PE over the interval 12:25 UT to 22:30 UT. A total of 754 spectra were acquired, covering the flare onset with PE, impulsive phase, and the extended post-eruption evolution. The spectral data were processed to correct for dark current, spectral curvature, flat-field variations, and background emission, following standard procedures described in previous studies (\citealt{RameshR2024_NewRes, JSingh2025_VELC, MPriyal2025_NearSun}). Figure~\ref{fig_spec_prof} shows spectral images before and after corrections at two different times with an example of fitting emission line with a gaussian function. This fitting to the emission line profile at each spatial position along the slit, gives key parameters: peak intensity ($I_p$), line width ($lw$; also known as FWHM after removing instrumental broadening), and doppler velocity ($V_d=\frac{\Delta \lambda}{\lambda_0}c$, where $\lambda_0=5302.8$~\AA~). The $V_d$ was computed relative to the mean line centroid measured in the background corona, which serves as the reference wavelength.

\begin{figure*}[!ht]
    \centering
    \includegraphics[width=0.8\linewidth]{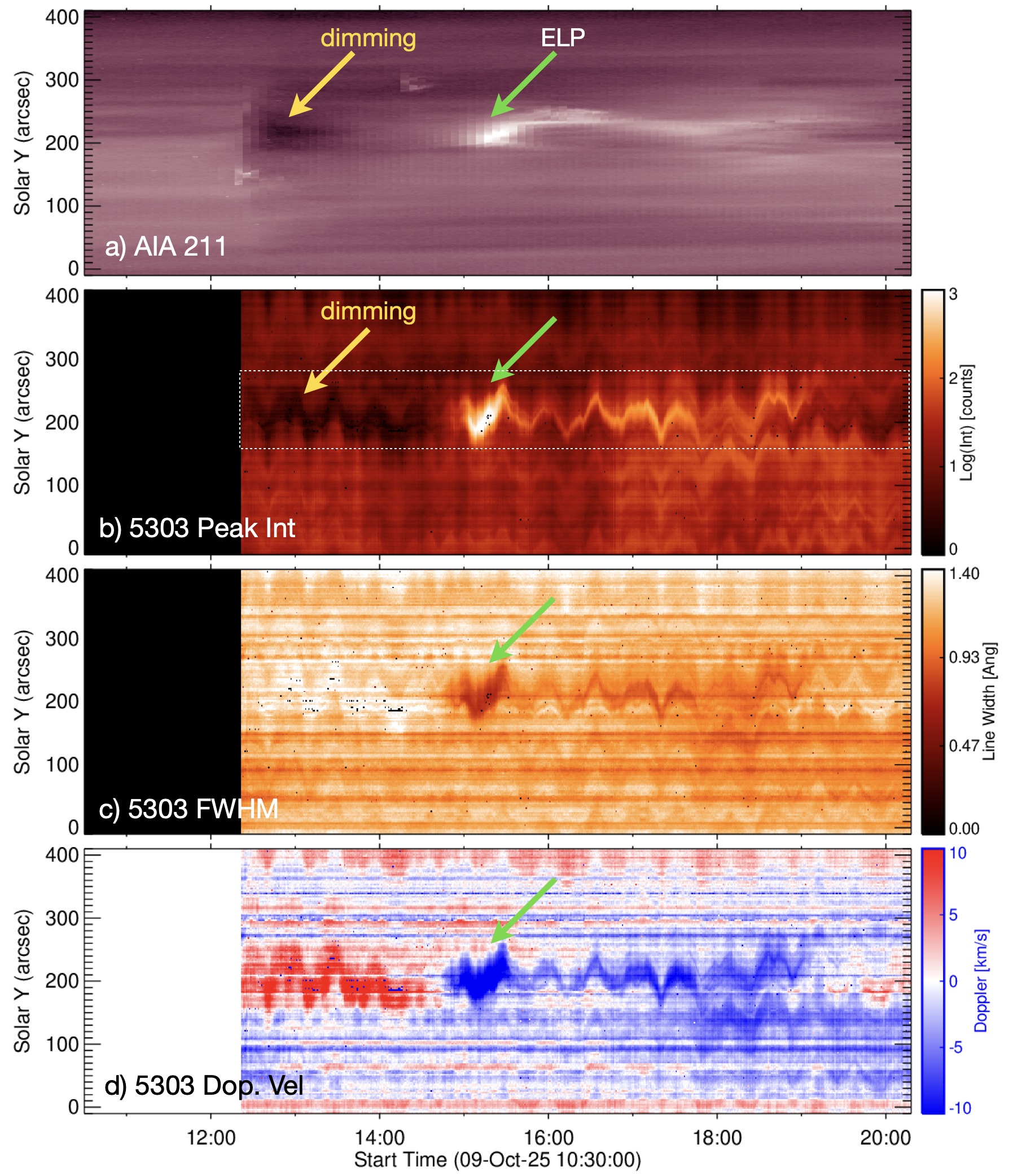}
    \caption{Stacked space-time map of slit at $X=1024$ arcsec. a) AIA 211 \AA~showing the dimming due to prominence eruption and brightened band linked to post-flare emission; b) Peak intensity in 5303 \AA~capturing dimming and gradual phase emission. Dotted rectangular box refer to the strip for y-spatial average. c) The corresponding line width and, d) Doppler velocity  }
    \label{fig_sl211_5303}
\end{figure*}

Figure~\ref{fig_sl211_5303} shows stacked space-time maps of peak intensity, line width, and Doppler velocity. Because the AIA 211~\AA~passband and 5303 \AA~arises from the Fe XIV with a similar temperature response (1.8-2 MK), for a comparison, the corresponding space–time map is constructed from AIA 211~\AA~observations by placing a virtual slit at $X=1024$ arcsec. The prominence-related emission appears as a horizontal band spanning $Y\approx170-280$ arcsec in all panels. The emission in the visible 5303~\AA~line closely tracks that of the AIA 211~\AA~channel, consistent with their similar temperature response around $\approx$2 MK. Following the onset of the PE at 12:10 UT, clear dimming signatures emerge in the 211~\AA~channel from about 12:25 UT onward, reflecting plasma evacuation along with the dense prominence material. A comparable behavior is also observed in AIA 193~\AA, which is sensitive to coronal plasma near $\approx$2 MK and similarly captures dimming at CME launching sites. This dimming phase is followed by a gradual recovery and enhanced emission, peaking at $\approx$15:24 UT, associated with post-eruption flare activity. The 5303~\AA~line emission exhibits a remarkable one-to-one correspondence with the AIA 211~\AA~evolution: a dimming phase persisting until ~14:40 UT, followed by a pronounced intensity increase. In particular, the peak emission at 15:24 UT aligns closely with the AIA 211~\AA~response, as well as with the average light curves of AIA 171 and 211~\AA~shown in Figure~\ref{fig_lgtcrv}. The oscillatory patterns are instrumental artifacts caused by spacecraft drift and pointing instabilities, for which no reliable correction method is currently available. The $lw$ corresponding to dimming period have higher values (1.5~\AA), which reduces to 0.7~\AA~during the post flare phase. Similarly, the $V_d$ show a net positive and negative velocities (upto 10 km\,s$^{-1}$) during these phases, that are likely associated with the prominence rise and the subsequent downward motion of plasma, provided projection effects with respect to limb direction.
\begin{figure*}
    \centering
    \includegraphics[width=0.8\linewidth]{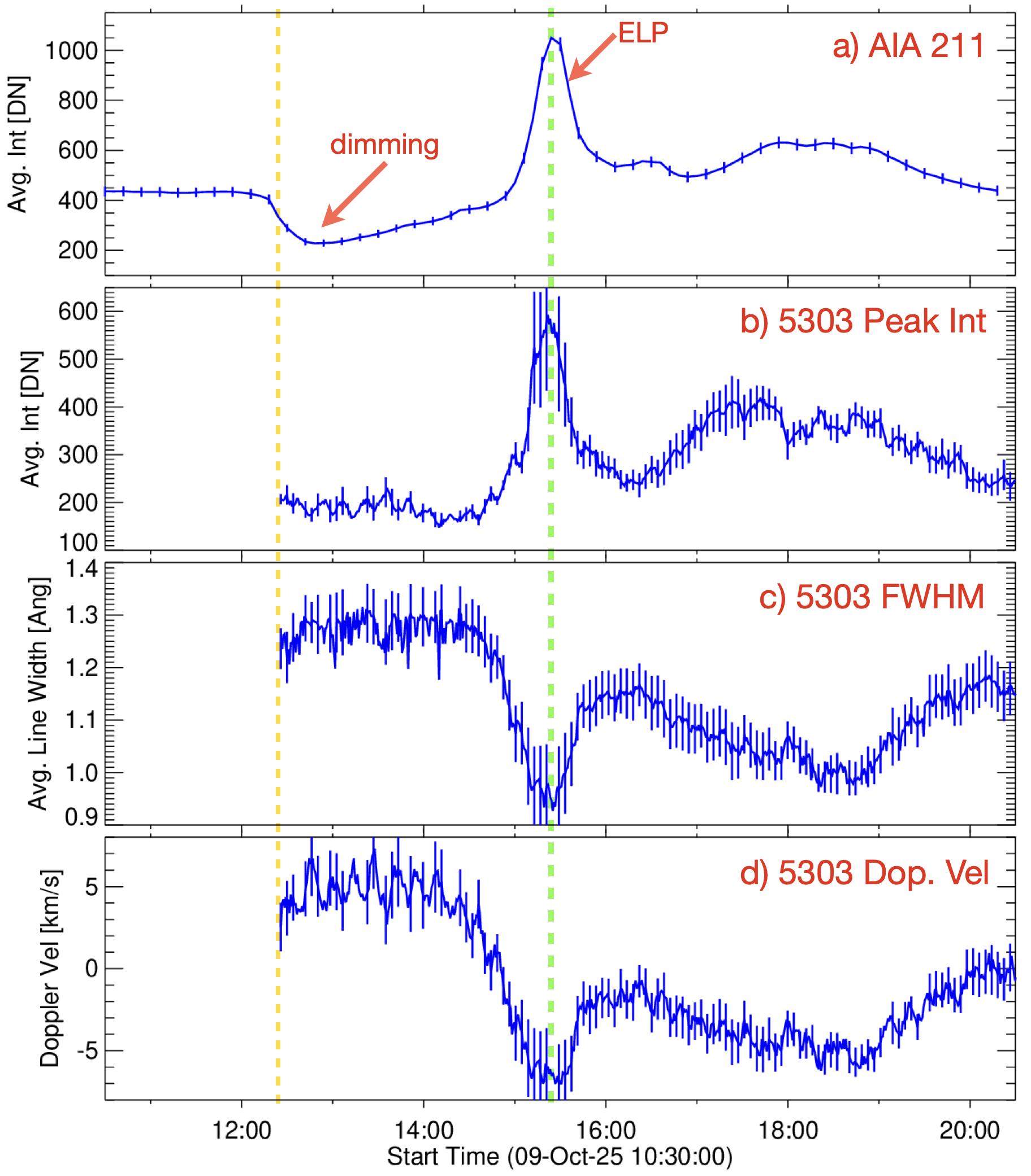}
    \caption{Averaged time profiles of a) AIA 211 \AA~ intensity, with error bars referring to mean error of averaging pixels in the virtual slit, b) VELC 5303 peak intensity, c) 5303 line-width (FWHM), and d) Doppler velocity of plasma in 5303~\AA.  Error bars refer to standard deviation of data points along the slit between $Y=170''$ and $280"$. Peak emission in AIA 211~\AA~ co-temporal with peak emission in 5303~\AA. Positive (negative) Doppler velocities likely indicate up flows (down flows) associated with the eruption of the prominence structure. }
    \label{fig_tprof_corr}
\end{figure*}
By averaging in the Y (spatial) direction of the slit between $Y=170''$ to 280'' (dotted box in Figure~\ref{fig_sl211_5303}b),  the temporal profiles are plotted in Figure~\ref{fig_tprof_corr}.  Averaging over this $Y$-strip also mitigates the effects of the oscillatory patterns, enabling a more reliable comparison with the AIA emission, and facilitating a more meaningful physical interpretation. A clear dimming signature is evident, and the averaged intensity profiles of 211~\AA~and 5303~\AA~agree well with the co-temporal peak emission at 15:25 UT. The peak line emission in 5303~\AA~increased by 65\%, while the AIA 211~\AA~intensity increased by 55\% relative to the pre-flare (before 12:11 UT) values. In addition to thermal motions, flare plasmas exhibit highly turbulent flows \citep{Ruan2023_MHD_Turb_flares} with excess line widths. Consequently, the measured $lw$ are as high as $1.28\pm0.06$~\AA~during the main flare phase (12:11-13:30 UT) and decreases to $0.92\pm0.07$~\AA~in the post-flare phase (13:30 UT onwards), following a recovery to $1.15\pm0.05$~\AA. The turbulent motions causing the excess $lw$ are linked to non-thermal velocity \citep{Stores2021_Turbulence_in_Flares}, which can be estimated by removing thermal component as $v_{nth}= {\left[{\left(\frac{lw}{\lambda }\right)}^{2}\frac{{{c}}^{2}}{4\mathrm{ln}(2)}-\left(\frac{2{{k}}_{{\rm{B}}}{T}}{{m}}\right)\right]}^{1/2}$, where $k_b$ is Boltzmann constant $k_b=1.38\times10^{-23}$ $\mathrm{J\,K^{-1}}$, mass of the Fe XIV ion $m=9.33\times10^{-26}$ Kg, and temperature of ions $T=1.8\times10^6$ K. With this expression, we found that the $v_{nth}$ decreases from $36.24\pm2.39$ km s$^{-1}$ during the impulsive flare phase to a lower value of $21.07\pm3.40$ km s$^{-1}$ during the gradual phase, and later shows a slight recovery to about $31.53\pm2.90$ km s$^{-1}$. These values are nearly 1.5 times as high as those measured during flare-less CMEs \citep{RameshR2024_NewRes,MPriyal2025_NearSun}. Similarly, $V_d$ exhibits a corresponding evolution, varying from $5\pm0.8$ km s$^{-1}$ during the impulsive phase to $-6.8\pm1.5$ km s$^{-1}$ during the gradual phase.

More importantly, owing to greeline correspondence with warm ELP emission, the $lw$ information becomes crucial for understanding the origin of the ELP emission, which remains under debate. Some studies suggest that the ELP occurs from additional energy injection and heating \citep{Woods2011_EVEObs_Flares, DaiY2013_EUV_LatePhase}, whereas others favor a prolonged cooling process following the main-phase reconnection heating \citep{LiuK2013_ELP, ZhangXueFei2022_Comp_AIA_5303}. During the ELP phase, beginning around 14:00 UT, the spatially averaged line width along the slit (panel~\ref{fig_sl211_5303}(c)) decreases predominantly until the peak emission observed in the AIA 211, 193, and 171 \AA~light curves. This decrease is 28\% in about 1.5 hours. As noted above, the $lw$ provides a measure of the non-thermal energy input during this period; therefore, its systematic decrease broadly supports a cooling process for the coronal loops after main phase reconnection. In contrast, if substantial energy injection through magnetic reconnection were responsible for the ELP, an enhancement in the $lw$ would be expected. Several studies have suggested that the additional heating associated with the ELP may be sufficient to produce warm plasma emission \citep{DaiY2013_EUV_LatePhase, Woods2011_EVEObs_Flares, ZhongY2021_EUV_LatePhase}, which is not strong enough to generate significant non-thermal line broadening. Nevertheless, observations of line-width variations in several ELP events need to study to establish statistically robust constraints on the underlying mechanism. 

It is worth noting that the $lw$ of the Fe XIV 5303~\AA emission line at its peak formation temperature of 1.8 MK is approximately 0.68~\AA\ \citep{Contesse2004_nonther_vel}. Statistical studies also report comparable values \citep{Kim2000_ElipSpec}. The minimum $lw$ values observed during the peak ELP phase (Figure~\ref{fig_sl211_5303}) are also close to this value. Therefore, once the plasma cools below the formation temperature of Fe XIV, the $lw$ is expected to decrease below $\sim$0.68~\AA, at which point further narrowing may no longer be reliably measurable from the 5303 \AA~emission line.

\section{Differential Emission Measure}
\label{DemAna}

\begin{figure*}[!ht]
    \centering
    \includegraphics[width=0.7\linewidth]{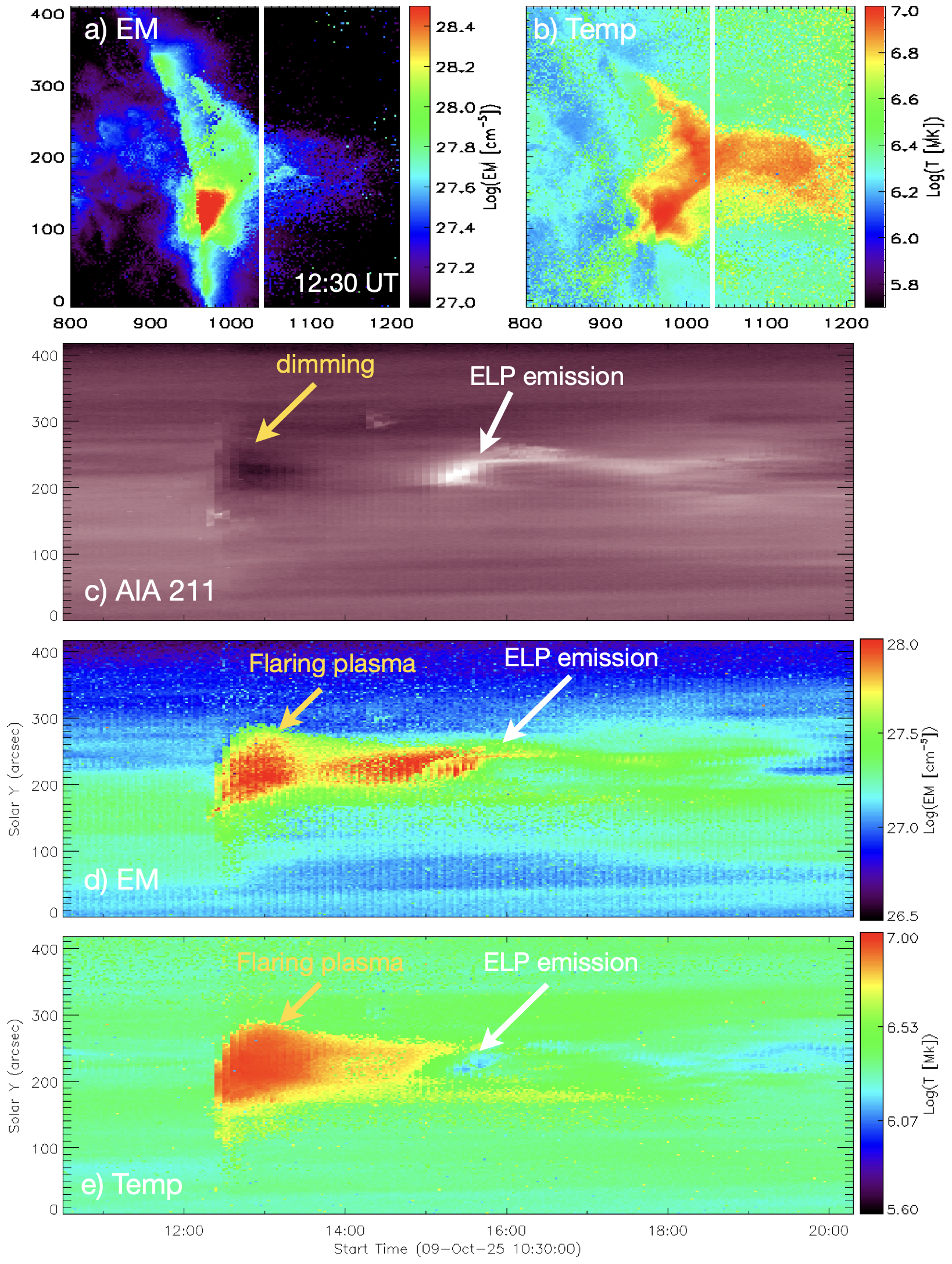}
    \caption{a-b) Maps of EM and DEM weighted temperature ($\bar{T}$) during flare phase. Vertical line refers to slit position at $X=1024''$. During flare, the prominence body is heated upto 10 MK by the reconnection. Axes are in arcsecond units in heliocentric coordinate frame.  c-e) Stacked space-time map of AIA 211 \AA~, emission measure, Temperature. Dimming corresponds to impulsive flare phase behind the erupting prominence and is characterized by emission in higher temperature (10 MK) channels. The ELP is characterized by emission in low temperature ($\sim2$ MK) channels.  }
    \label{fig_st_dem}
\end{figure*}

Using the six EUV channels of AIA, we further investigated the thermal properties of the coronal plasma through differential emission measure (DEM) analysis. For each passband, the DEM was reconstructed on a pixel-by-pixel basis using the forward-fitting routine \texttt{xrt\_dem\_iterative2.pro} available in Solar SoftWare (SSW; \citealt{freeland1998}), employing the corresponding temperature response functions and observed EUV intensities. Following previous studies \citep{Chengx2012_dem, Vemareddy2017_PromEru}, from DEM at each pixel, maps of the emission measure ($ EM=\int DEM(T) dT $) and DEM-weighted average temperature ($\bar{T}=\frac{\int DEM(T) T dt}{\int DEM(T)dt}$) are computed within the temperature range of $5.6<LogT<7.3$. These maps are displayed in Figure~\ref{fig_st_dem}.

The EM and temperature ($\bar{T}$) maps reveal a clear contrast between the prominence structure and the surrounding background corona in both EM and $\bar{T}$. Prior to the eruption, the EM and $\bar{T}$ are of the order of $10^{27}$ cm$^{-5}$ and $10^{6}$ K, respectively. With the onset of the eruption and flare reconnection, both quantities increase by nearly an order of magnitude, as shown in the top panels of Figure~\ref{fig_st_dem} indicating substantial plasma heating associated with magnetic reconnection. A pronounced height-dependent gradient in both EM and temperature is also evident throughout the erupting structure. During the post-eruption phase, around 15:30 UT, both EM and $\bar{T}$ decrease relative to their peak values during the flare phase, reflecting the gradual cooling of the coronal plasma.

To investigate the temporal evolution of these thermal properties, space--time maps of EM and $\bar{T}$ were constructed using a slit positioned at $X=1024''$ and compared with the corresponding AIA 211~\AA\ observations (Figure~\ref{fig_st_dem}(c-e)). In the strip $Y=170-280''$, the flaring plasma is quite hot and reaches temperatures exceeding 10 MK during the interval 12:25--13:40 UT, during which AIA 211 \AA\ show with dimming and subsequently cools as the event progresses into the gradual phase. It is important to note that from approximately 14:00 UT onward, the EM increases to above $1.5\times10^{28}$ cm$^{-5}$, while the corresponding temperature decreases below 4 MK, indicating the continued cooling of the flaring plasma following the main flare phase. In particular, the ELP emission in AIA 211~\AA\ (indicated by the arrow) is associated with EM values of $\sim10^{28}$ cm$^{-5}$, while the derived plasma temperature decreases to $\sim1$--2 MK. This further supports the interpretation that prolonged plasma cooling dominates the ELP phase, which is primarily manifested in the AIA channels sensitive to plasma at temperatures of $\sim1$--3 MK, as shown in Figure~\ref{fig_lgtcrv}. Consequently, the enhanced intensity observed in AIA 211~\AA, and correspondingly in the Fe XIV 5303~\AA\ green line, is primarily associated with warm coronal plasma at temperatures of approximately $\sim2$ MK.

\begin{figure*}[!ht]
    \centering
    \includegraphics[width=0.8\linewidth]{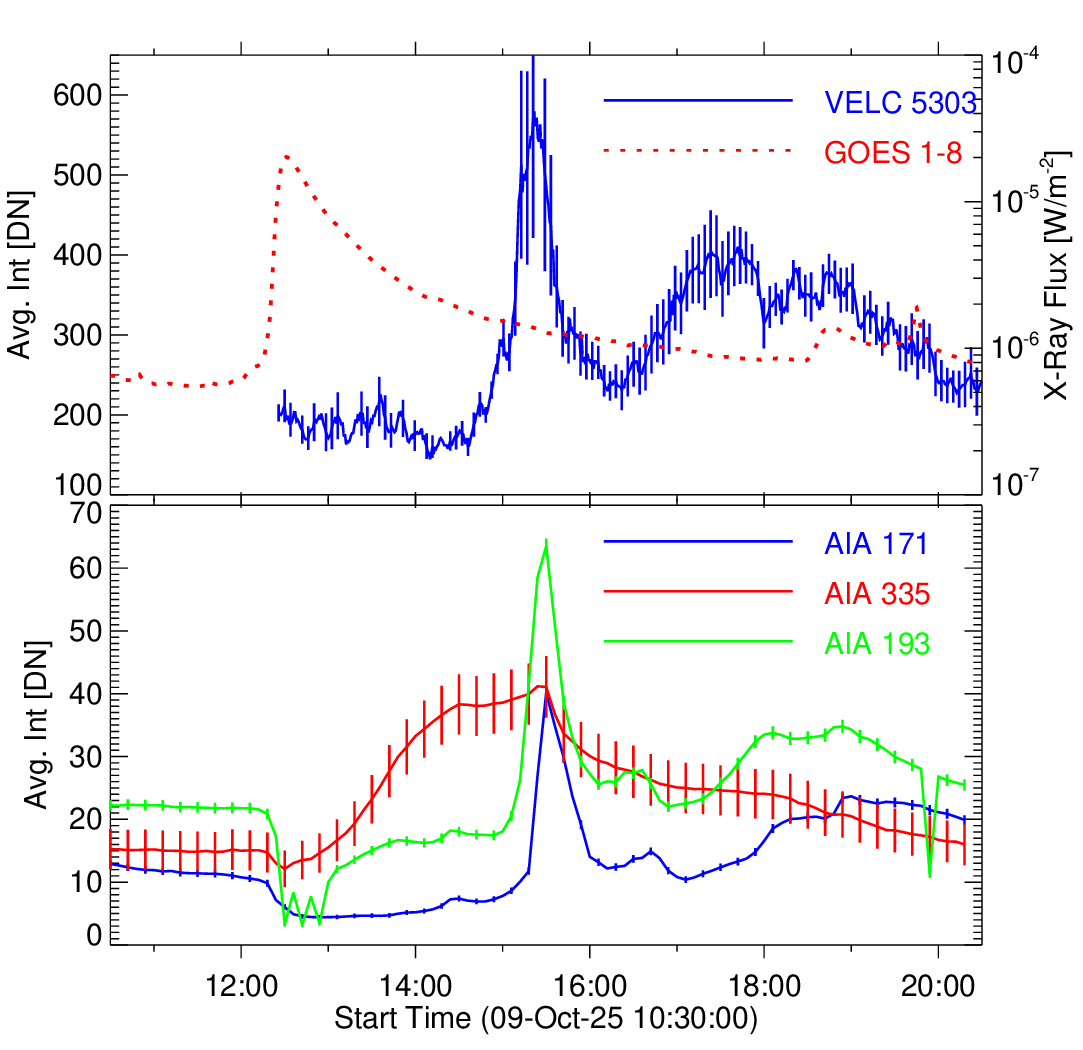}
    \caption{ Averaged time profiles deduced from the slit ($Y=170''-280''$) of AIA 335, 171, 193 \AA~in comparison with VELC 5303 \AA~.  The ELP peak emission is co-temporal with 5303 \AA, which is 173 minute delayed from X-ray flare peak (red dotted line).   
 }
\label{fig_euv_tp}
\end{figure*}

This thermal evolution provides a natural explanation for the observed behavior of the 5303~\AA\ emission. During the main flare phase, the average plasma temperature is significantly higher ($\sim3.7$ MK ) than the peak formation temperature of the Fe XIV 5303~\AA\ line, resulting in weak or absent green-line emission. As the plasma cools during the gradual phase, the average temperature decreases to approximately $2\times10^{6}$ K, close to the peak formation temperature of Fe XIV. Consequently, the conditions become favorable for enhanced 5303~\AA\ emission, explaining its strong appearance during the ELP phase and its relative absence during the main flare phase.

\section{Summary and Discussion}
\label{Summ}
We investigated the PE that occurred on 9 October 2025 on the west limb of the Sun, associated with an M2.0 flare that started at 12:11 UT. Fortunately, the VELC sit-and-stare spectroscopic observations captured this event from 12:25 UT onward, enabling us to study the thermal properties of the flaring plasma. The slit was positioned at 1.06 R$_\odot$ ($\approx42$ Mm) in the corona, sampling the erupting prominence with spectral information in the 5303~\AA~emission line. During a flare, the impulsive phase is associated with rapid heating and subsequent evaporation of hot plasma driven by thermal conduction and/or nonthermal electron bombardment, whereas the gradual phase is dominated by radiative cooling \citep{Antiochos1980_Radiative_cooling, Cargill1995_Cooling_Sol_FlarePlasma}. In addition to the X-ray peak, in the gradual phase the EUV emissions exhibit secondary peaks in different passbands that appear sequentially in order of decreasing temperature, corresponding to the cooling of the flaring plasma \citep{Chamberlin2012_ThermEvol_Flares}. This phenomenon is now recognized as ELP emission \citep{Woods2011_EVEObs_Flares,ZhongY2021_EUV_LatePhase,YaWang2026_ELP}, originating from extended coronal loop systems and spectroscopic observations from VELC offer valuable insight for understanding it more thoroughly.

Our detailed analysis revealed an enhanced emission in the 5303~\AA~during the ELP phase, with a peak at 15:24 UT that coincides precisely with the AIA 211~\AA~EUV emission.  This ELP peak also co-existed with AIA 171, 193, 335 \AA~wavebands as shown in Figure~\ref{fig_euv_tp} and separated by 173 minutes. Such a prominent peak may not be seen in eruption cases without flares \citep{Vemareddy2017_PromEru}. The ELP was attributed to secondary phase of heating that takes place well after the heating of main flare loops \citep{Woods2011_EVEObs_Flares,ZhongY2021_EUV_LatePhase, YaWang2026_ELP}. However, this delayed heating should be much weaker compared to the main flare heating, therefore mainly enhancing the emissions at intermediate temperatures. Moreover, the AIA 211, 335, and 193~\AA\ passbands have peak temperature responses near 2.0 MK, 2.5 MK, and 1.3 MK, respectively \citep{lemen2012}, making them sensitive to plasma temperatures comparable to the peak formation temperature of the Fe XIV 5303~\AA\ green line. Therefore, the observed correspondence between the ELP emission in these channels and the enhanced 5303~\AA\ emission suggests that the ELP may also encompass coronal green-line emission. A similar inference was drawn in a recent study of \citet{ZhangXueFei2022_Comp_AIA_5303} who compared EUV and 5303 intensity images noting  a high spatial correlation of coronal features. However, this intriguing possibility needs to be confirmed by analyzing additional events, then only the green line emission can be regarded as a proxy for EUV emission during the gradual phase of flares.


Green line spectral data also sheds light on the line width and Doppler velocity information which further indicate dynamic plasma evolution during the flare. Enhanced line widths during the impulsive phase suggest strong turbulent and non-thermal plasma motions, which decrease during the gradual phase. The high-cadence temporal observations provide unprecedented insight into the line-width variations during dynamic events, offering a perspective that is fundamentally different from earlier studies, which primarily focused on height-dependent variations \citep{Mierla2008_linewidth}. For the first time, the $line-width$ information uncover the the dominant process responsible for ELP emission. The line-widths decreases gradually upto the peak ELP emission, which is broadly consistent with the cooling scenario of coronal loops \citep{LiuK2013_ELP, YaWang2026_ELP}. The delayed energy injection by reconnection is weak, causing ELP in warm channels without enhancing the line-widths. However, the line width studies during more ELP flare events is required to substantiate this inference.

From the line widths, the estimated non-thermal velocities vary from about 36 km s$^{-1}$ during the impulsive phase to 21 km s$^{-1}$ during the gradual phase which are within the range reported by \citet{Doschek2014_PlasmaDyn}. These values are nearly 1.5 times as large as those observed in flare-less CMEs \citep{RameshR2024_NewRes, Mpriyal2024_DataProc}. Doppler velocities also evolve systematically, showing positive velocities during the eruptive rise phase and negative velocities during the later cooling and downflow phase, likely reflecting upward and downward plasma motions associated with the PE. 

Flare-induced enhancements in EUV irradiance can significantly influence the Earth’s ionosphere and thermosphere (e.g., \citealt{Donnelly1978_}). The impact of ELP on the global total electron content (TEC), and consequently on Earth's near-space environment, has also been investigated (e.g., \citealt{Bekker2024_Flare_TEC}). For the X9.3-class flare on 6 September 2017, \citet{LiuXuanquing2024_ELP_Ionosp} reported that the enhancement in TEC during the ELP was greater than that during the main flare phase. Their study further showed that the ELP produced measurable enhancements in the ionospheric electric field and equatorial electrojet, consistent with observational evidence. In this perspective, the correspondence of ELP emission in greenline opens up the possibility of investigating the impact of enhanced EUV irradiance using ground/space-based coronagraphs equipped with green-line spectroscopy by providing valuable diagnostics of the thermal and dynamical properties of the flaring plasma.

With continuous long-duration sit-and-stare spectroscopic observations and EUV images, this study provides the first observational evidence of a correspondence of coronal green-line emission from flaring plasma with the ELP emission. The availability of regular space-based spectroscopic observations will increase the likelihood of capturing similar events, enabling detailed investigations of coronal plasma evolution and its role in a wide range of dynamic solar processes.  

%
\begin{acknowledgments}
We thank P. Savarimuthu, E. Yuvashree, and S. Nagashree
for their help in processing the VELC data. The VELC payload team members are sincerely appreciated for their tireless efforts in developing the payload. ADITYA-L1 is an observatory class mission, funded and operated by the Indian Space Research Organization (ISRO). Data obtained with the different payloads on board ADITYA-L1 are archived at the \href{https://pradan.issdc.gov.in/al1/protected/browse.xhtml}{Indian Space Science Data Centre (ISSDC)} of ISRO. SDO is a mission of NASA's Living With a Star Program. We thank the reviewers for critical comments and suggestions.
\end{acknowledgments}
%
%
\bibliographystyle{aasjournalv7}
%

\end{document}